\documentclass[manuscript]{aastex}

\slugcomment{Not to appear in Nonlearned J., 45.}

\shorttitle{R-Process Yields} \shortauthors{Li et al.}

\begin{document}

\title{Astrophysical Origins For The Unusual Chemical Abundance Of The Globular Cluster
Palomar 1}


\author{Ping Niu\altaffilmark{1,2}, Wenyuan Cui\altaffilmark{1},
Bo Zhang\altaffilmark{1} and Geying Xie\altaffilmark{3}}

\affil{1. Department of Physics, Hebei Normal University, No.20
East of South 2nd Ring Road, Shijiazhuang 050024, China \\
2. Department of Physics, Shijiazhuang  University, Shijiazhuang
050035, China \\
3. Department of Mathematics and Physics, Hebei Institute of
Architecture Civil Engineering, Zhangjiakou 075024, China}

\altaffiltext{3}{Corresponding author. E-mail address:
zhangbo@mail.hebtu.edu.cn}

\begin{abstract}

Observed unusual elemental abundances in globular cluster Palomar 1
(Pal 1) could provide important information for us to study the
relation between the globular cluster and our Galaxy. In this work,
we study the abundances of $\alpha$ elements, Fe-peak elements and
neutron-capture elements in Pal 1. We found that the abundances of
the SNe Ia and main s-process components of Pal 1 are larger than
those of the disk stars and the abundances of the primary component
of Pal 1 are smaller than those of the disk stars with similar
metallicity. The Fe abundances of Pal 1 and the disk stars mainly
originate from the SNe Ia and the primary component, respectively.
Although the $\alpha$ abundances dominantly produced by the primary
process for the disk stars and Pal 1, the contributions of the
primary component to Pal 1 are smaller than the corresponding
contributions to the disk stars. The Fe-peak elements V and Co
mainly originate from the primary and secondary components for the
disk stars and Pal 1, but the contributions of the massive stars to
Pal 1 are lower than those of the massive stars to the disk stars.
The Y abundances mainly originate from the weak r-component for the
disk stars. However the contributions of the main s- and main
r-components to Y are close to those of the weak r-component for Pal
1. The Ba abundances of Pal 1 and the disk stars mainly originate
from the main s-component and the main r-component, respectively.
Our calculated results imply that the unusual abundances of Pal
could be explained by the top-light IMF for Pal 1's
progenitor-system.

\end{abstract}


\section{Introduction}

The heavy elements (Z$>$30) are mainly produced by neutron-capture
processes \citep{Bur57}. The slow neutron-capture process
(s-process) contains two categories. The weak s-process mainly
produces the lighter neutron-capture elements and takes place in the
massive stars \citep{Lam77,Rai91,The00}. The main s-process occurs
in the AGB stars and is directly responsible for the heavier
elements \citep{Bus99}. The rapid neutron-capture process
(r-process) also contains two categories. The main r-precess could
occur in SNe II with 8-10 $M_{\odot}$ progenitor and mainly produces
the heavier elements \citep{Cow91,Sne08}. However the weak r-process
might take place in the SNe II with 11-25 $M_{\odot}$ progenitors.
The SNe II primarily produce the weak r-process elements and eject
the light and Fe-peak  elements \citep{Tra04}.

Globular clusters are the living fossils to study the Galactic
evolution history since their extremely ancient age. Historically
Palomar 1 (Pal 1) was thought as a globular cluster, since its
location is higher than the Galactic plane \citep{Har96}. However,
the red giants in Pal 1, whose averaged metallicity [Fe/H] is about
-0.6, show so metal-rich for a globular cluster of the outer halo
\citep{Ros98b}. Based on the study of the color-magnitude diagram of
Pal 1, \cite{Ros98a} suggested its age lies the range of 6.3-8 Gyr.
In this case, Pal 1 was regarded as the youngest globular cluster.
Recently, using high-resolution spectra, \cite{Sak11} derived
chemical abundances for four red giants in Pal 1. They found that
element abundances of Pal 1 are unusual comparing to those of the
disk stars. The abundances of Mg, Si, Ca, Ti, Co, V and Y of Pal 1
are lower than those of the disk stars with similar metalliciy.
However the abundances of Ba and La are higher than those of the
disk stars. \cite{Sak11} suggested that Pal 1 is not the traditional
GC and should be accreted from a dwarf galaxy by our Galaxy. The
unusual abundances should imply that Pal 1 should have an unusual
formation environment and history.

Nearby galaxies should have complex formation histories and chemical
evolution process \citep{Tol09}. They should have a close relation
to our galaxy. For example, some stars and clusters in the Galaxy
should originate from the dwarf galaxies. In this case, these stars
retain the chemical signatures of its progenitor system. The
chemical signatures, such as lower [$\alpha$/Fe] ratios, and star
formation histories of nearby dwarf galaxies have been studied for
many years \citep{Tol03,Lan08,Tsu11}. Historically, the explanation
of low [$\alpha$/Fe] ratios in dwarf galaxies is that the light
elements are deficient, since a low star formation rate (SFR) and
extra Fe from SNe Ia \citep{Tin79}. Based on the detailed abundance
analysis, \cite{McW13} concluded that the $\alpha$-element
deficiencies in the Sagittarius (Sgr) dwarf spheroidal galaxy result
from an top-light initial mass function (IMF), which is relatively
deficient in the highest mass stars. \cite{Wei13,Kro13} studied the
effects of galaxy mass on stellar IMF. They found that the
integrated galaxy IMF varies from top-light to top-heavy in
dependence of galaxy type. The top-light IMF should be associated to
the dwarf galaxies because the galaxies have not sufficient massive
clouds to form the highest-mass stars \citep{Oey11}. On the other
hand, the IMFs seem to vary among more massive early type galaxies
\citep{Van12}. \cite{Con12} found that more massive early type
galaxies are associated to bottom-heavy IMF, because many early
supernovae drive up the supersonic turbulence in the giant clouds,
which leads to formation of more low mass stars.

It is important to note that the abundances of individual stars of
Pal 1 present the accumulated effects from the time of the
progenitor-system formed to the time of Pal 1 formed. For purpose of
investigating the complex formation mechanism and history of Pal 1,
it is necessary to study the elemental abundances completely,
containing light elements, Fe-peak elements and heavy elements. In
this paper, using the abundance approach given by \cite{Li13b}, we
investigated the astrophysical origins of chemical abundances of Pal
1 and the disk stars. The results and discussions are given in
Section 2. Our conclusions are presented in Section 3.

\section{Results And Discussions}

It is indicated that the stellar abundances could not be explained
by single astrophysical reason \citep{All06}. For
exploring the origins of light and heavy elements in Pal 1, we adopt
the abundance approach from \cite{Li13b}. The abundance of element i
should be express as:

\begin{equation}
N_{i}(Z) = (C_{r,m}N_{i,r,m} + C_{pri}N_{i,pri} + C_{s,m}N_{i,s,m} +
C_{sec}N_{i,sec}+ C_{Ia}N_{i,Ia}) \times 10^{[Fe/H]},
\end{equation}

where $N_{i,r,m}$, $N_{i,s,m}$, $N_{i,pri}$, $N_{i,sec}$ and
$N_{i,Ia}$ are the abundances of the main r-, the main s-, the
primary, the secondary process and SNe Ia, respectively, which have
been scaled to the corresponding abundances of Solar system.
$C_{r,m}$, $C_{s,m}$, $C_{pri}$, $C_{sec}$ and $C_{Ia}$ are the
corresponding component coefficients. The abundance $N_{i}$ is the
number of atoms of element i which is scaled to $10^{6}$ Si atoms of
the solar abundances.

The primary light elements and Fe-peak elements are produced in the
massive stars ($M\geq10 M\odot$) during hydrostatic burning and the
weak r-process elements were produced in the SNe II with progenitor
mass $M\geq10 M\odot$. Because the primary light elements, Fe-peak
elements and the weak r-process elements are ejected from the
massive stars, these elements could be combined as one component,
which is called the ``primary component". On the other hand, the
secondary light elements and Fe-peak elements are produced in the
massive stars ($M\geq10 M\odot$) during hydrostatic burning and the
weak s-process elements are produced in the massive stars during
core He burning and shell C burning. The yields of weak s-process
also have the secondary nature. Because the secondary light
elements, Fe-peak elements and the weak s-process elements come from
the massive stars, these elements could be combined as ``secondary
component" \citep{Li13b}. Note that the abundances of secondary
elements are observable only for higher metallicity, since their
yields decrease with decreasing metallicity. In this case, the
primary component $N_{i,pri}$ contains the abundances of primary
elements and the weak r-elements. The secondary component
$N_{i,sec}$ contains the abundances of the secondary elements and
the weak s-elements. Both primary and secondary components are
produced in the massive stars. The abundances of the secondary
elements and the weak s-elements are adopted from \cite{Li13b} and
\cite{Rai93} respectively. The abundances $N_{i,r,m}$ and
$N_{i,pri}$ are adopted from \cite{Li13a}. The abundances $N_{i,Ia}$
are adopted from \cite{Tim95} and updated the Fe, Cu, and Zn values
from \cite{Mis02}. \cite{Kob98} reported that the SNe Ia events
could not occur for the progenitors with low metallicity
[Fe/H]$<-1.0$. This implies that the SNe Ia events only occur on the
higher metalllicy. In this work, we suppose that the abundance
pattern produced by SNe Ia do not vary with metallicity, because the
yields of SNe Ia depend on metallicity weakly \citep{Iwa99}. In this
case, the adopted abundance pattern produced by SNe Ia can be
thought as the average abundance pattern produced by SNe Ia.

From the observations, \cite{Sak11} reported that compared to the
disk stars, the [Ba/Y] values are apparently high, which indicates
that the Pal 1 stars were contaminated by the low-metallicity and
low-mass AGB stars. In these AGB stars, the s-process prefers
heavier s-nuclei rather than lighter s-nuclei, because an iron-seed
nucleus can capture more neutrons \citep{Bus01}. Obviously, the
different abundance patterns between Pal 1 and the disk stars
attribute to the different chemical characteristics between Pal 1
and the disk stars. The abundances $N_{i,s,m}$ in Equation (1) are
adopted from the results with [Fe/H]=-0.6 presented by \cite{Bus01}
for Pal 1. By comparison, for the disk stars, the abundance pattern
$N_{i,s,m}$ in Equation (1) are adopted from the results with
[Fe/H]=-0.6 calculated by \cite{Tra99} (see their Figures 6$-$12).

Equation (1) contains five components. The corresponding component
coefficients can be derived by looking for the minimum $\chi^{2}$.
Adopting the average abundances in Pal 1 stars \citep{Sak11} and the
average abundances of the disk stars with [Fe/H]=-0.6 \citep{Red06},
we can obtain the best-fit coefficients. For the abundances of the
Solar system, $C_{r,m}\simeq C_{pri}\simeq C_{s,m}\simeq
C_{sec}\simeq C_{Ia}\simeq 1$. Furthermore, we can compare the
component coefficients of Pal 1 with the corresponding coefficients
of the disk stars to investigate the astrophysical reasons of
unusual abundances in Pal 1 stars.

The calculated results for Pal 1 and the disk stars are shown in
Figure 1 (a) and (b). In top panels, the filled circles indicate the
observed abundances and the solid lines represent the fitted
results. In bottom panels, the individual relative offsets
($\bigtriangleup log \varepsilon = log \varepsilon(cal)-log
\varepsilon(obs)$) and standard calculated errors for Pal 1
and the disk stars are shown, respectively. It is confirmed
from the figure 1 (a) and (b) that the calculated results
is validity.

In Figure 2, the values of the component coefficients $C_{r,m}$,
$C_{pri}$, $C_{s,m}$, $C_{sec}$, $C_{Ia}$ and the associated
errors are displayed for Pal 1 and the disk stars, respectively.
From the figure we can see that the values of $C_{s,m}$, and
$C_{Ia}$ of Pal 1 are higher than those of the disk stars, which
implies that the contributions from the two components to the
abundances of Pal 1 are larger than those of the disk stars. On the
other hand, the value of $C_{pri}$ of Pal 1 is lower than that of
the disk stars, which implies that the contribution of the component
to Pal 1 are smaller than that to the disk stars. Furthermore, the
values of $C_{sec}$ and $C_{r,m}$,of Pal 1 are close to those of the
disk stars. The abundances of each component can be derived
quantitatively using corresponding component coefficient.

Since the various categories of elements are produced in different
astrophysical sites, the element abundances of Pal 1 should be
directly effected by IMF of its progenitors' system. We can study
the characters of the IMF through the contributions of different
processes to the element abundance of Pal 1. The primary component
and secondary component are produced by the massive stars with
M$\gtrsim10M_{\odot}$. On the other hand, the main r-process should
take place in the SNe II with 8$\thicksim10 M_{\odot}$ progenitors
\citep{Cow91,Sne08} and the main s-process mainly occurs in the
1.5$\thicksim3M_{\odot}$ AGB stars \citep{Bus99}. Based on the
calculated results, the coefficients ratios of Pal 1 to the disk
stars for primary process, secondary process, main r-process, SNe Ia
and main s-process are about 0.6, 1.0, 1.2, 3.0 and 4.5
respectively. The ratios have a trend that increase with decreasing
progenitor mass, which should be an evidence that the IMF of Pal 1's
progenitor-system (e.g., dwarf galaxy in which cluster Pal 1 formed)
is top-lighter than that of the Galaxy. Note that the coefficient
ratios of Pal 1 to the disk stars are determined by their IMFs
dominantly and the coefficient ratios should depend on their SFRs
weakly. If the IMF of Pal 1 is similar to that of the Galaxy, the
coefficient ratios of Pal 1 to the disk stars should be close to a
constant. So the unusual abundances of Pal 1 are mainly due to the
top-light IMF of its progenitor system. Recently, the signatures of
the top-light IMF in the elemental abundances of the Fornax dwarf
galaxy  and the Sagittarius dwarf galaxy have been reported
\citep{Li13b,McW13}. The results imply that the IMF of
Pal 1's progenitor-system have the similar characters with that of
the dwarf spheroidal galaxies mentioned above.

Our approach is based upon the measured abundances of the sample
stars, thus the associated uncertainties would be included in the
calculations. Taking the component coefficient of main s-process of
Pal 1 as an example, we discuss the uncertainties of the
coefficients. Adopting $C_{r,m}$=3.5, $C_{pri}$=1.2, $C_{sec}$=1.1
and $C_{Ia}$=0.9, the top panel in Figure 3 shows the calculated
ratios [Ba/H] as a function of the component coefficients $C_{s,m}$.
There is only a range of the component coefficients in Figure 3,
$C_{s,m}=1.3^{+1.6}_{-0.4}$, in which the calculated ratios fall
into observed limits of [Ba/H]. The bottom panel in Figure 3
illustrates the reduced $x^{2}$ values are the function of the component
coefficients $C_{s,m}$. There is a minimum of $x^{2}=0.58$ at
$C_{s,m}$=1.3 with a $1\sigma$ error bar, which indicates
that the component coefficient is constrained well. Adopting this
approach, we derived the errors of other component coefficients:
$C_{r,m}=3.5^{+1.5}_{-1.2}$, $C_{pri}=1.2^{+0.2}_{-0.9}$,
$C_{sec}=1.1^{+0.3}_{-1.1}$ and $C_{Ia}=0.9^{+0.3}_{-0.2} $for Pal
1. $C_{s,m}=0.3^{+0.1}_{-0.2}$ $C_{r,m}=2.9^{+0.7}_{-0.3}$,
$C_{pri}=2.1^{+0.3}_{-0.5}$, $C_{sec}=1.1^{+0.8}_{-0.7}$ and
$C_{Ia}=0.3^{+0.3}_{-0.1}$ for the disk stars. The calculated errors
of component coefficients are shown in Figure 2. In this case,
the calculated abundance errors should be close to the observational
errors, since the component coefficients are restrained by the
observational errors. The average observed errors of all elements
for the Pal 1 stars and the disk stars are about 0.17 and 0.07 dex,
respectively \citep{Sak11,Red06}. In this work, for simplicity, we
take the average observed errors as the standard calculated errors
of elemental abundances for the two category stars.

Because different elements in Pal 1 came from different
astrophysical sites on different epoches, the abundances of
individual stars present the accumulated effects from the time of
the progenitor-system formed to the time of Pal 1 formed. In order
to reveal the astrophysical reasons of unusual abundances of Pal 1,
we show the component abundance ratios of Fe, Mg, Co, Y and Ba in
Pal 1 and compare them with the disk stars in Figures 4-8. In the
figures, the filled pentagons represent the calculated abundance ratios and
the associated error bars represent the calculated errors. From
Figure 4, we know that for the disk stars, the Fe element mainly
produced by primary process. Although the observed Fe abundances of
the disk stars are similar to those of Pal 1, the abundances of the
primary and SNe Ia components of the disk stars are obviously
different from those of Pal 1. For the Pal 1, the Fe abundance
mainly originate from the SNe Ia. The calculated results imply that
the primary Fe abundances of Pal 1 are lower than those of the disk
stars, which agree in a top-lighter IMF of Pal 1's
progenitor-system.

\cite{Sak11} reported that the observed Mg abundances of Pal 1 are
lower than those of the disk stars by about 0.4 dex. From Figure 5
we know that for the disk stars and Pal 1, although the Mg
abundances dominantly produced by the primary process, the primary
Mg abundances of Pal 1 are smaller than those of the disk stars.
This should mean that the proportion of the massive stars in Pal 1's
progenitor-system is smaller than that of the massive stars in the
Galaxy. The observed unusual abundances of other $\alpha$ elements
in Pal 1 could be also explained by the top-light IMF of Pal 1's
progenitor-system.

\cite{Sak11} showed that abundances of some Fe-peak elements, such
as V and Co of Pal 1 are smaller than those of the disk stars. Taken
Co as an example, Figure 6 shows that, for the disk stars and Pal 1,
although the Co abundances mainly originate from the primary and
secondary components of the massive stars, the contributions of the
massive stars to Pal 1 is lower than those of the massive stars to
the disk stars, which also agree in the top-light IMF of Pal 1's
progenitor-system. The lower abundance of V in Pal 1 could be
explained by the similar reason.

It is indicated that the Y abundances mainly originate from the weak
r-component for the disk stars in Figure 6. However, for pal 1, the
Y abundance originate from three components: main r-, weak r- and
main-s components in the same figure. \cite{Sak11} have found that
the Y abundances of Pal 1 are lower than those of the disk stars
about 0.5 dex. From the figure we know that the main reason of the
lower Y abundance is that the abundance of weak r-component of Pal 1
is smaller than those of the disk stars. The abundances of s-process
component for Pal 1 should be mainly contaminated by the low
metallicity AGB stars with 1.5-3 $M_{\odot}$. In these AGB stars,
the s-process prefers heavier s-nuclei (e.g., Ba, La) rather than
lighter s-nuclei (e.g., Sr, Y), because an iron-seed nucleus can
capture more neutrons \citep{Bus01}. This should be another reason
of the lower Y abundances in Pal 1. As Figure 7 shows, for element
Y, the observed abundance including error bar falls into the
calculation error limits.

From Figure 8, we can see that the Ba abundances mainly originate
from the main r-component for the disk stars. On the other hand, for
Pal 1, the Ba abundance mainly originate from the main s-component.
\cite{Sak11} reported that the Ba abundances of Pal 1 are larger
than those of the disk stars. From the figure we know that the main
reason of higher Ba abundance is that the Ba abundance of main
s-component of Pal 1 is higher than those of the disk stars, which
should attribute to the top-light IMF for the Pal 1's
progenitor-system.

\section{CONCLUSIONS}

The element abundances of Pal 1 contained a great deal of
information of element nucleosynthesis and evolution history. In
this paper, using the abundance-decomposed approach, we explore the
astrophysical reasons of the unusual observed abundances in Pal 1
stars. We found that the component coefficients of main s-process
and SNe Ia of Pal 1 are higher than those of the disk stars.
However, the component coefficient of primary process of Pal 1 is
lower than that of the disk stars. The results should imply that the
IMF of Pal 1's progenitor-system is top-lighter than that of the
Galaxy.

The Fe abundances mainly produced by primary process for the disk
stars and the Fe abundances mainly originate from the SNe Ia for Pal
1 stars. Although the $\alpha$-element abundances dominantly
produced by the primary process for the disk stars and Pal 1, the
contribution of primary component of the massive stars to Pal 1 is
lower than those of the massive stars to the disk stars, which
should be the astrophysical reason of the observed low-$\alpha$
abundances in Pal 1. The Fe-peak elements V and Co mainly produced
in the massive stars for the disk stars and Pal 1, but the
contributions of the massive stars to Pal 1 is lower than those of
the massive stars to the disk stars, which might be the
astrophysical reason of the observed low abundances of V and Co in
Pal 1. That the Y abundances of weak r-component for Pal 1 are
smaller than that of weak r-component for the disk stars should be
the main reason of the low Y abundances in Pal 1' stars.
Furthermore, that the Ba abundances of the main s-component for Pal
1 are higher than that of main s-component for the disk stars is the
astrophysical reason of the high Ba abundance in Pal 1 stars. As a
whole, the observed unusual abundances in Pal 1'stars could be
explained by the top-light IMF for the Pal 1's progenitor-system.

Our results obtained from the abundance-decomposed approach can
provide more constraints on the astrophysical environment of Pal 1.
Overall, we hope our work can give a beneficial guidance for judging
chemical origin of Pal 1. Of cause, more observed elemental
abundances about the stars of Pal 1 are needed to investigate the
relation between Pal 1 and a dwarf galaxy.

\acknowledgments

We thank the anonymous referees for insightful comments which
improved this paper greatly. We also thank Dr Jianrong Shi for the
careful review of the manuscript and for fruitful discussion. This
work has been supported by the National Natural Science Foundation
of China under 11273011, U1231119, 10973006 and 11003002, the
Natural Science Foundation of Hebei Provincial Education Department
under grant Z2010168, XJPT002 of Shijiazhuang University, the
Natural Science Foundation of Hebei Province under Grant
A2011205102, A2011210017, and the Program for Excellent Innovative
Talents in University of Hebei Province under Grant CPRC034.

\clearpage

\begin{figure}[t]
 \centering
 \includegraphics[width=1.0\textwidth,height=0.4\textheight]{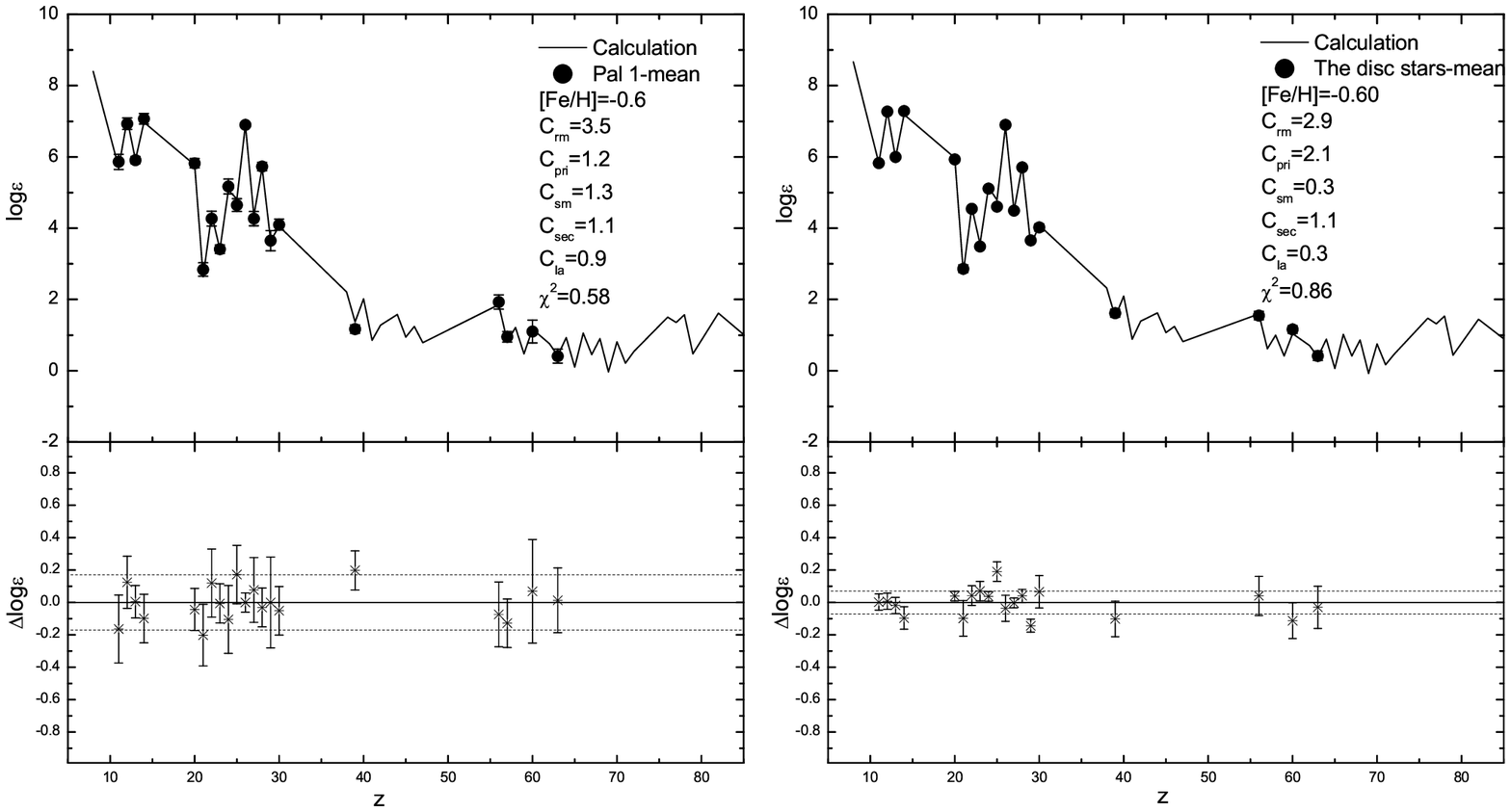}
\caption{Top panel: Best fitted results of Pal 1 and the disk stars.
The filled circles with error bars represents the observed
abundances, the solid lines are the calculated results. Bottom
panel: the individual relative offsets
($\Delta\log\varepsilon(X)\equiv\Delta\log\varepsilon(X)_{cal}-\Delta\log\varepsilon(X)_{obs}$)
and the standard calculated errors in $log \varepsilon$ (dashed lines) for Pal 1 and the disk stars.}
\end{figure}

\begin{figure}[t]
 \centering
 \includegraphics[width=1.0\textwidth,height=0.6\textheight]{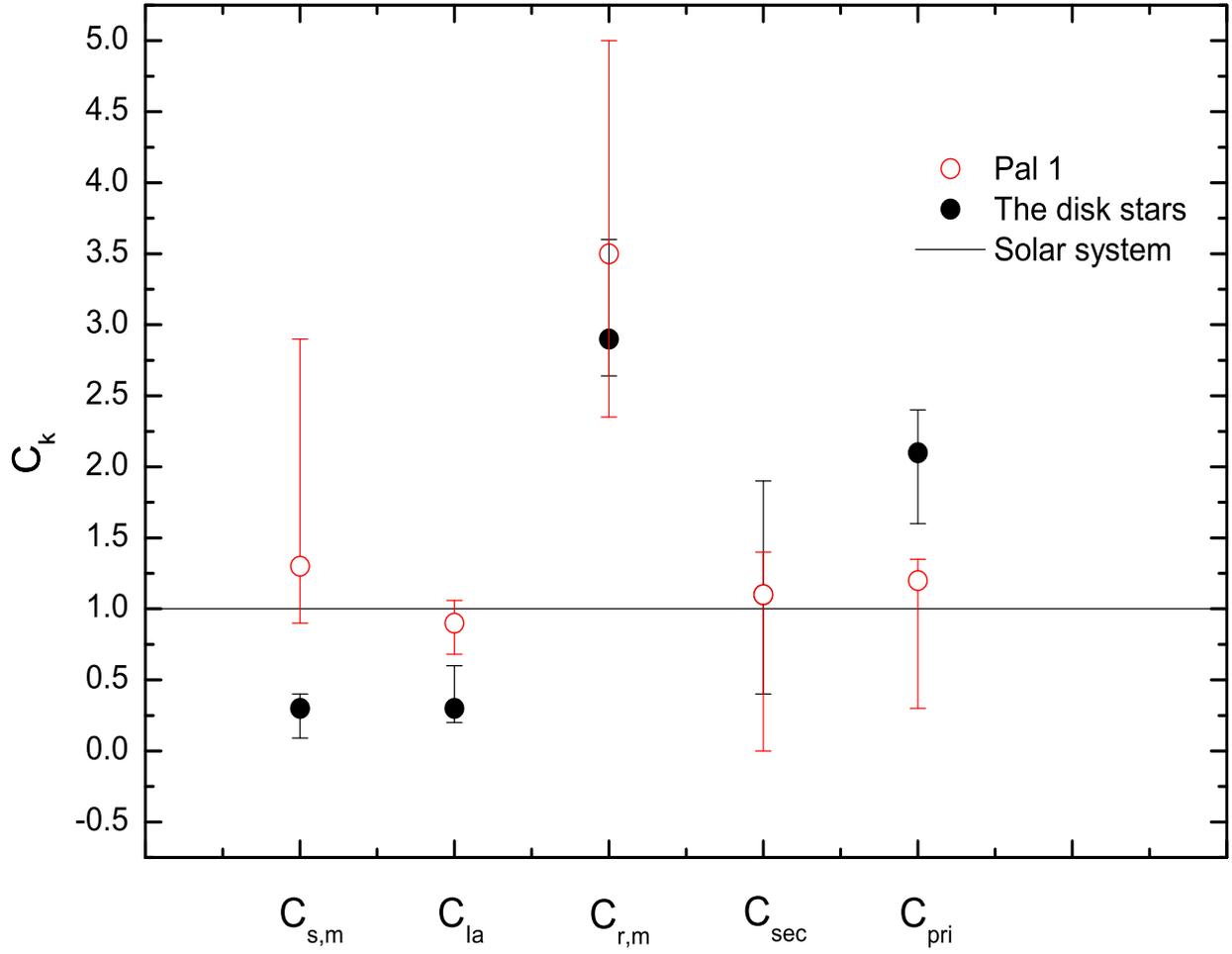}
\caption{The values of the component coefficients
 $C_{r,m}$, $C_{pri}$, $C_{s,m}$, $C_{sec}$, $C_{Ia}$ and the associated errors are
displayed for Pal 1 (open circles)and the disk stars (filled circles) respectively.}
\end{figure}

\begin{figure}[t]
 \centering
 \includegraphics[width=1.0\textwidth,height=0.6\textheight]{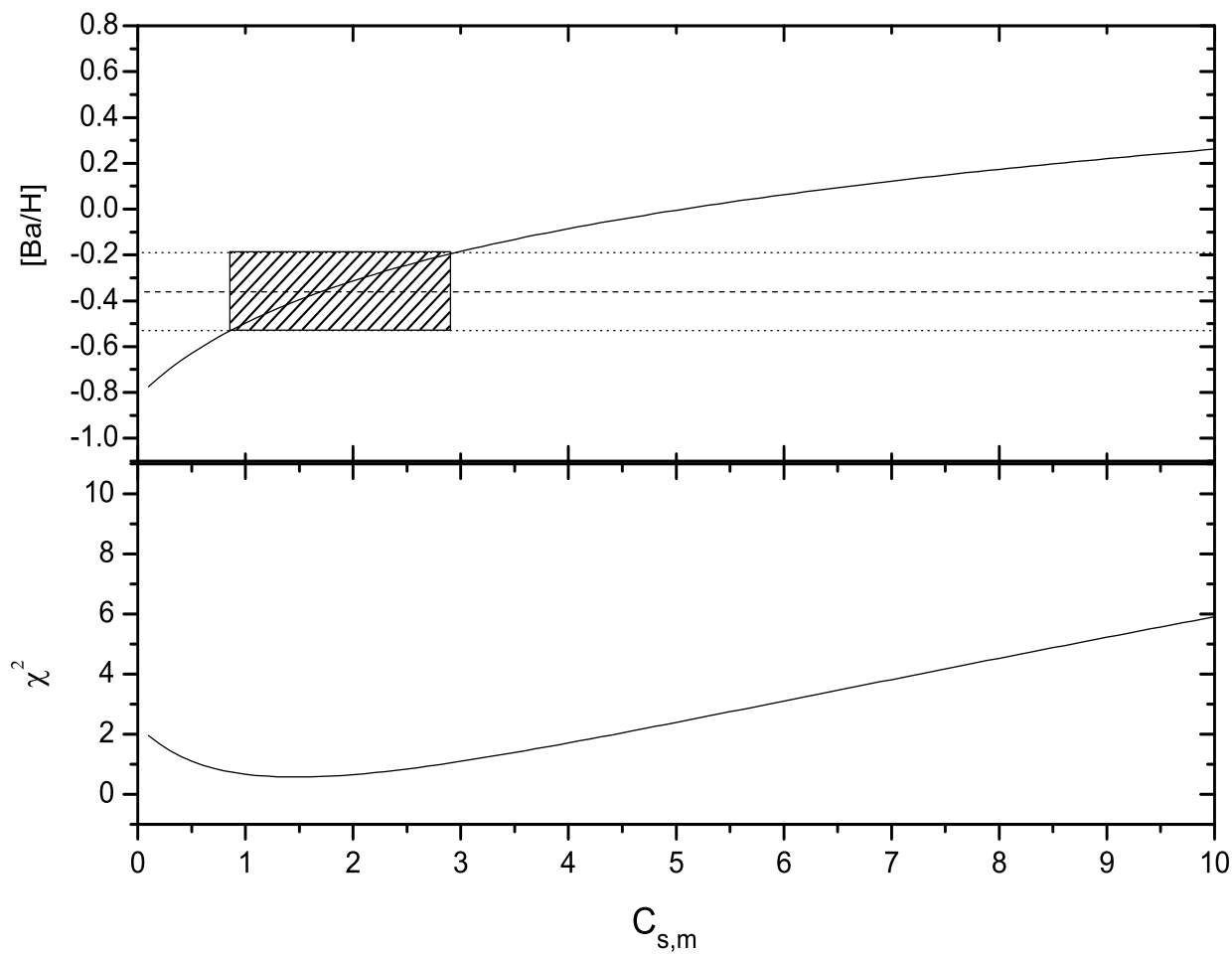}
\caption{: The calculated ratios [Ba/H] (top panel) and the reduced $x^{2}$ (bottom panel) as a function of
the component coefficients $C_{s,m}$, in a model with $C_{r,m}=3.5$, $C_{pri}=1.2$, $C_{sec}=1.1$ and $C_{Ia}=0.9$. Solid
curves represents the calculated results, and dashed horizontal lines refer to
the observed values, with errors plotted by dotted lines. The shaded
areas illustrate the allowed regions for the theoretical model.}
\end{figure}

\begin{figure}[t]
 \centering
 \includegraphics[width=1\textwidth,height=0.6\textheight]{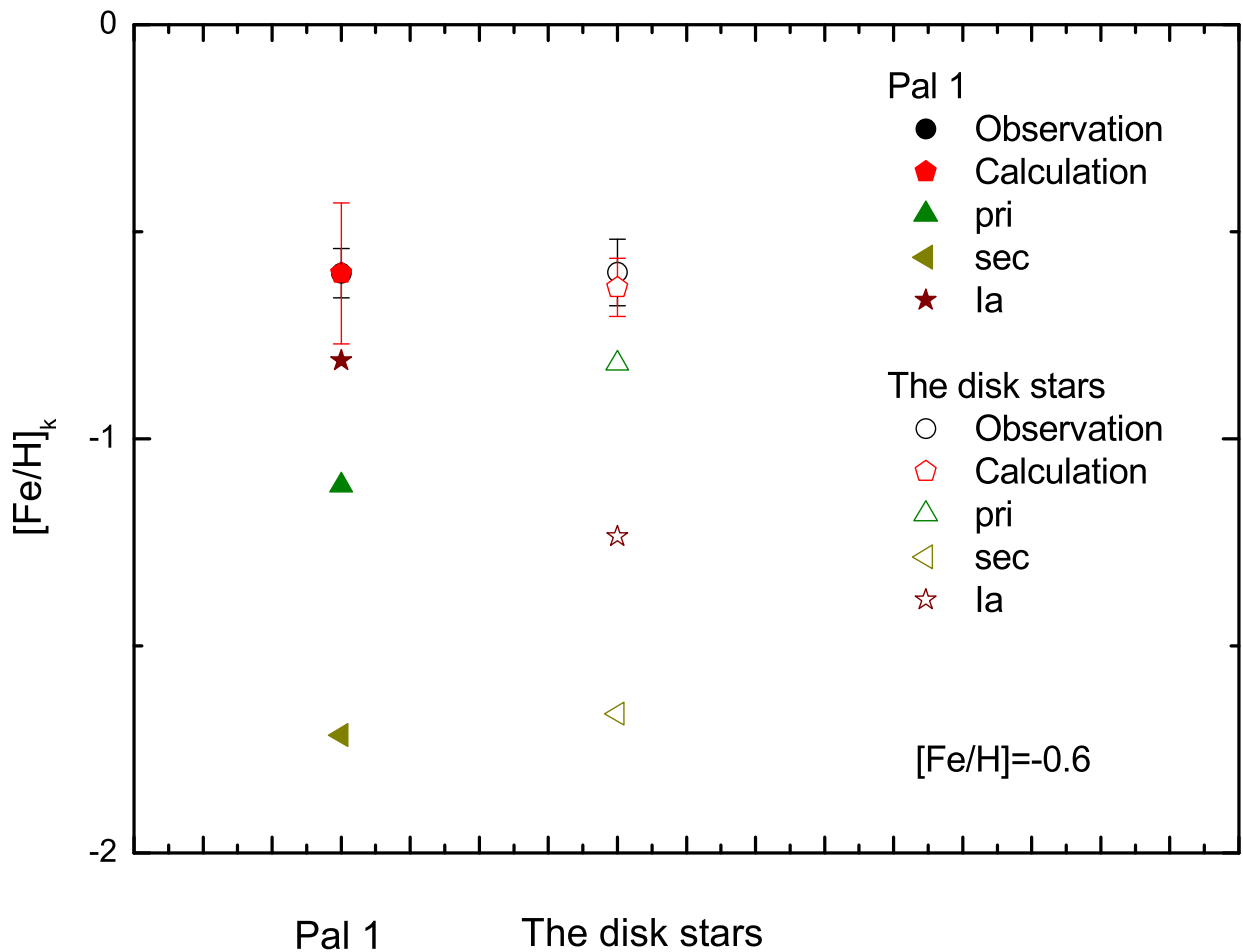}
\caption{The Component ratios of the individual processes with
metallicity [Fe/H]=-0.6 for element Fe in Pal 1 and the disk stars.
For Pal 1, the filled circles with error bars represent observed
abundance ratios and the associated error; the filled pentagons
with the error bars represent the calculated abundance ratios
and the associated error; The filled up triangles, filled left
triangles, filled stars, filled diamonds, and filled down triangles
represent the ratios of primary, secondary, SNe Ia,
main r- and main s-component, respectively. For the disk stars,
the open circles with error bars represent observed abundance ratios
and the associated error; the open pentagons with the error bars
represent the calculated abundance ratios and the associated error;
the open up triangles, open left triangles, open stars, open
diamonds and open down triangles represent the ratios
of primary, secondary, SNe Ia, main r- and main s-component, respectively.}
\end{figure}

\begin{figure}[t]
 \centering
 \includegraphics[width=1\textwidth,height=0.6\textheight]{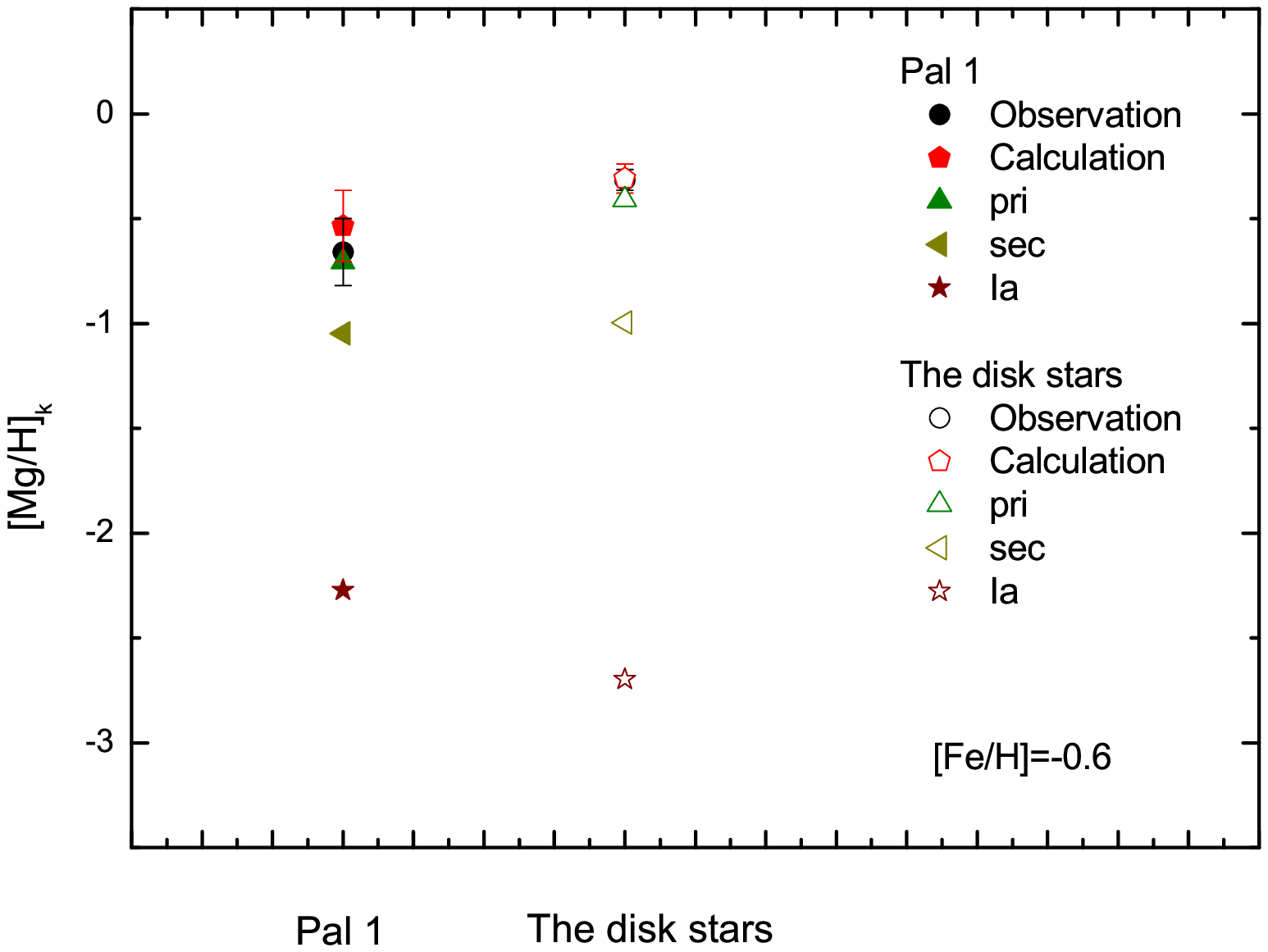}
\caption{Component ratios of the individual processes with
metallicity [Fe/H]=-0.6 for element Mg  in Pal 1 and the disk
stars. The meaning of symbols are the same as Figure 4.}
\end{figure}

\begin{figure}[t]
 \centering
 \includegraphics[width=1\textwidth,height=0.6\textheight]{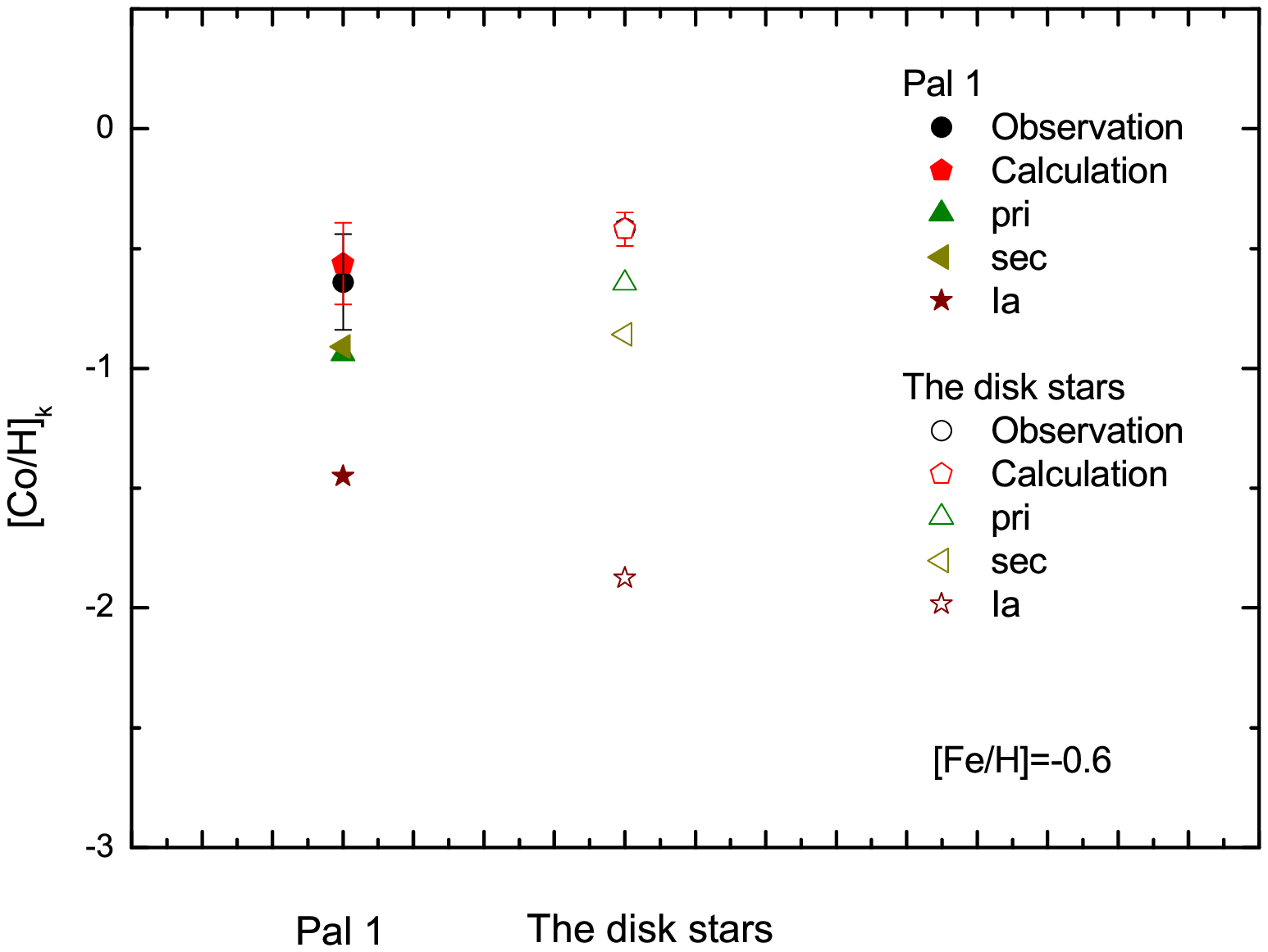}
\caption{Component ratios of the individual processes with
metallicity [Fe/H]=-0.6 for element Co  in Pal 1 and the disk
stars. The meaning of symbols are the same as Figure 4.}
\end{figure}

\begin{figure}[t]
 \centering
 \includegraphics[width=1\textwidth,height=0.6\textheight]{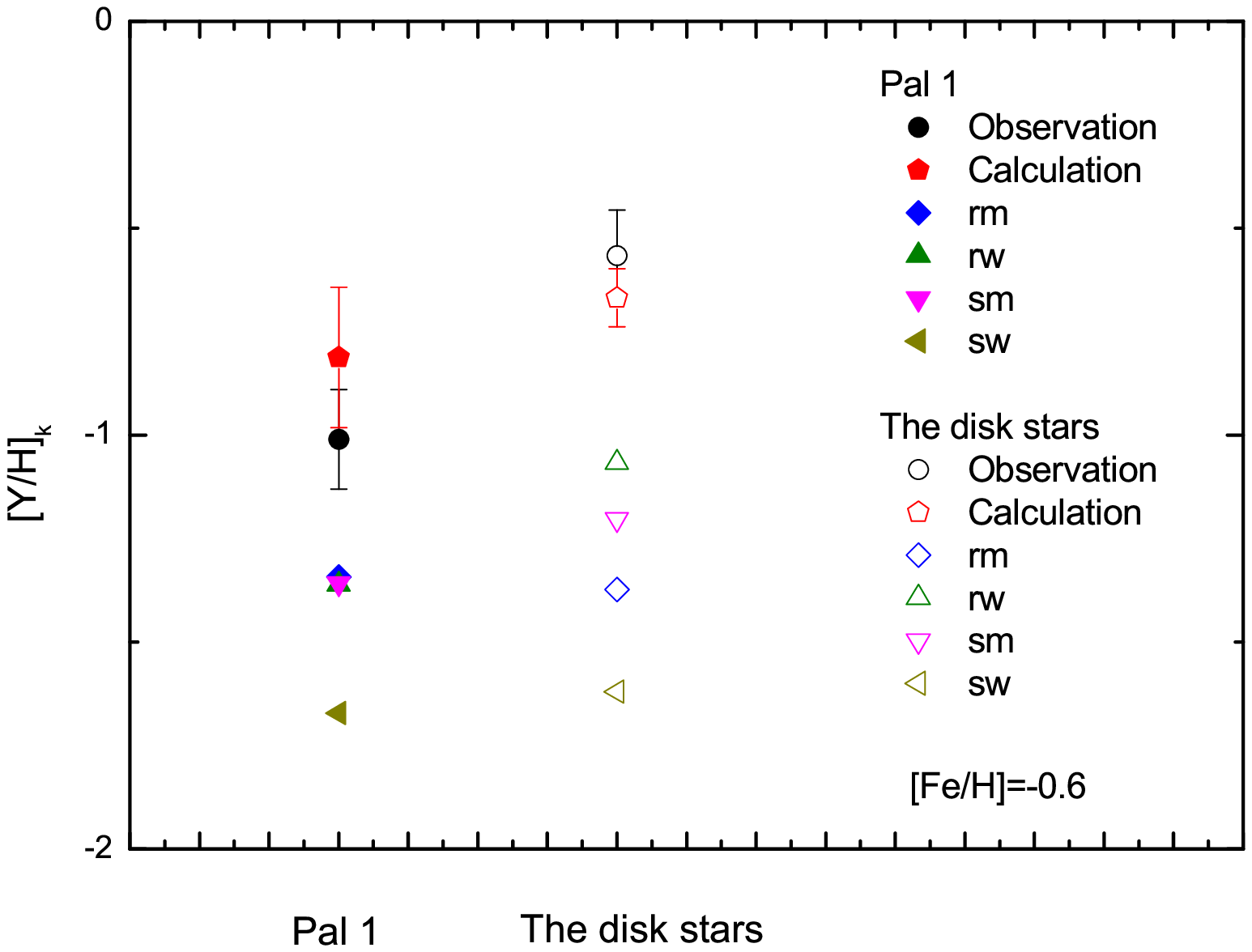}
\caption{Component ratios of the individual processes with
metallicity [Fe/H]=-0.6 for element Y  in Pal 1 and the disk stars.
The meaning of symbols are the same as Figure 4.}
\end{figure}

\begin{figure}[t]
 \centering
 \includegraphics[width=1\textwidth,height=0.6\textheight]{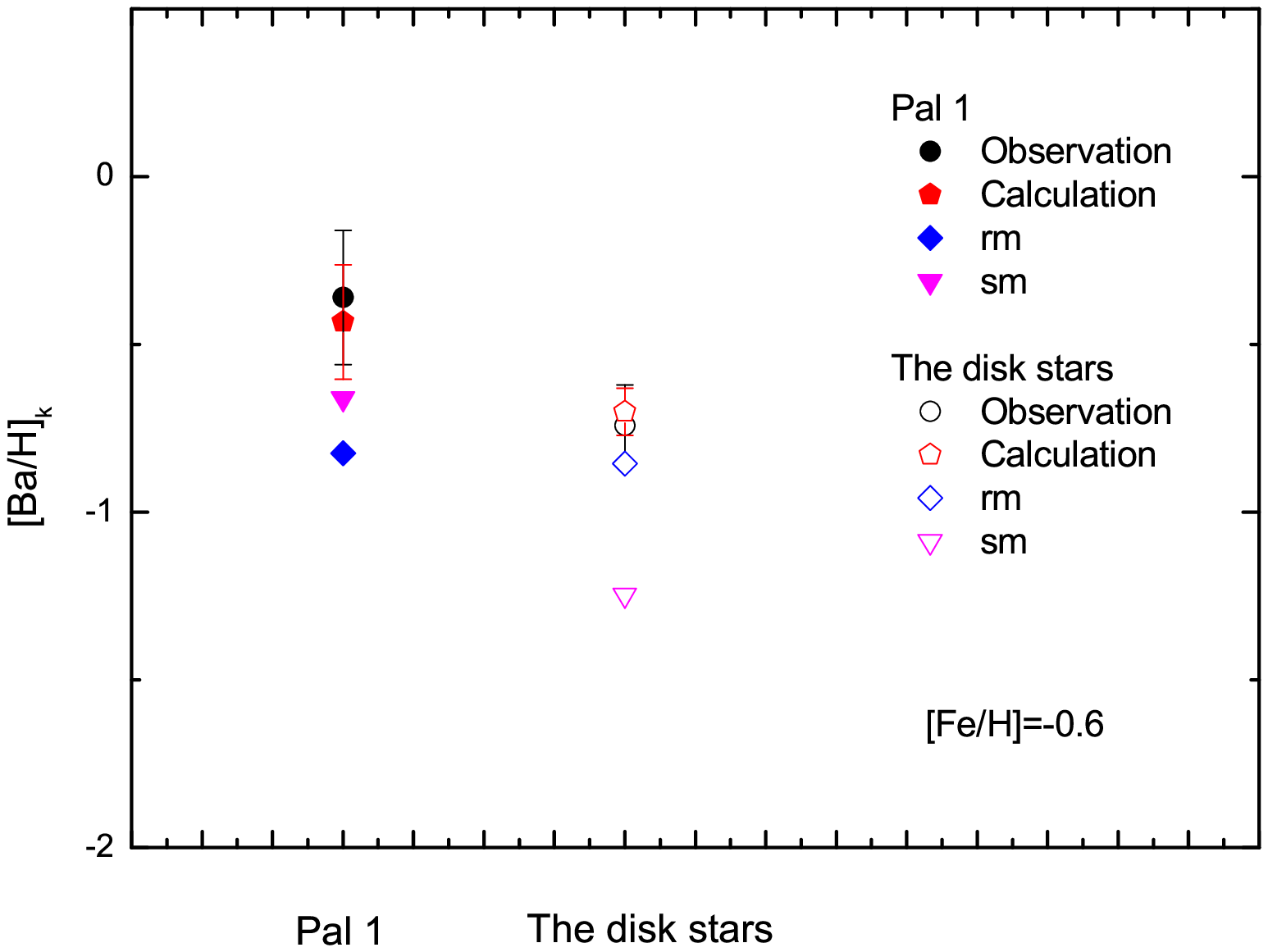}
\caption{Component ratios of the individual processes with
metallicity [Fe/H]=-0.6 for element Ba  in Pal 1 and the disk
stars. The meaning of symbols are the same as Figure 4.}
\end{figure}

\clearpage

\end{document}